\documentclass[a4paper,11pt]{article}
\usepackage{pos}

\usepackage{mathrsfs}
\usepackage{enumitem}
\usepackage{xifthen}
\DeclareMathAlphabet{\mathpzc}{OT1}{pzc}{m}{it}

\newcommand{\aprop}{\mathrel{\raise.3142ex\hbox{$\propto$}\kern-.76em\lower.8ex\hbox{$\approx$}}}
\newcommand{\simprop}{\mathrel{\raise.3142ex\hbox{$\propto$}\kern-.66em\lower.6ex\hbox{$\sim$}}}

\newcommand{\vka}{\ensuremath{\varkappa}}

\newcommand{\vfi}{\ensuremath{\varphi}}

\newcommand{\Eta}{\ensuremath{\mathrm{H}}}

\newcommand{\ionfrac}[2][env]{\ensuremath{\phi_{#2}^{\ifthenelse{\equal{#1}{}}{}{\scriptscriptstyle\langle\mathrm{#1}\rangle}}}}
\newcommand{\ionfracUp}[2][env]{\ensuremath{\Phi_{#2}^{\ifthenelse{\equal{#1}{}}{}{\scriptscriptstyle\langle\mathrm{#1}\rangle}}}}
\newcommand{\fraction}[2][env]{\ensuremath{f_{\ifthenelse{\equal{#2}{j}\OR\equal{#2}{k}}{#2}{\rm #2}}^{\ifthenelse{\equal{#1}{}}{}{\scriptscriptstyle\langle\mathrm{#1}\rangle}}}}
\newcommand{\numspec}[2][env]{\ensuremath{N_{\ifthenelse{\equal{#2}{j}\OR\equal{#2}{k}}{#2}{\rm #2}}^{\ifthenelse{\equal{#1}{}}{}{\scriptscriptstyle\langle\mathrm{#1}\rangle}}}}
\newcommand{\temperature}[2][env]{\ensuremath{T_{\ifthenelse{\equal{#2}{j}\OR\equal{#2}{k}}{#2}{\rm #2}}^{\ifthenelse{\equal{#1}{}}{}{\scriptscriptstyle\langle\mathrm{#1}\rangle}}}}
\newcommand{\density}[2][env]{\ensuremath{n_{\ifthenelse{\equal{#2}{j}\OR\equal{#2}{k}}{#2}{\rm #2}}^{\ifthenelse{\equal{#1}{}}{}{\scriptscriptstyle\langle\mathrm{#1}\rangle}}}}

\newcommand{\abundance}[2][]{\ensuremath{\vfi_{\ifthenelse{\equal{#2}{j}\OR\equal{#2}{k}}{#2}{\scriptscriptstyle\rm #2}}^{\ifthenelse{\equal{#1}{}}{}{\scriptscriptstyle\langle\mathrm{#1}\rangle}}}}
\newcommand{\Enhance}[2][]{\ensuremath{\Eta_{\ifthenelse{\equal{#2}{j}\OR\equal{#2}{k}}{#2}{\scriptscriptstyle\rm #2}}^{\ifthenelse{\equal{#1}{}}{}{\scriptscriptstyle\langle\mathrm{#1}\rangle}}}}

\newcommand{\maxval}[1]{\ensuremath{\widehat{#1}}}

\newcommand{\av}[2][env]{\ensuremath{\left\langle#2\right\rangle}^{\ifthenelse{\equal{#1}{}}{}{\!\scriptscriptstyle\langle\mathrm{#1}\rangle}}}
\newcommand{\Prob}[3][]{\ensuremath{\mathscr{P}^{#3}_{\!\mathrm{\scriptscriptstyle #1}#2\,}}}

\newcommand{\cdiff}[3][]{\ensuremath{\mathscr{D}^{#3}_{\mathrm{#1}#2\,}}}

\newcommand{\tscale}[3][]{\ensuremath{\tau^{#3}_{\mathrm{#1}#2}}}

\newcommand{\area}[3][]{\ensuremath{\mathscr{A}^{#3}_{\mathrm{#1}#2}}}

\newcommand{\pinj}{\ensuremath{p_{\rm inj}}}
\newcommand{\xinj}{\ensuremath{\zeta_{\rm inj}}}
\newcommand{\adindex}[2][\langle env\rangle]{\ensuremath{\gamma_{\rm ad#2}^{\scriptscriptstyle\rm#1}}}

\newcommand{\bsh}{\ensuremath{\beta_{\rm sh}}}

\newcommand{\down}{\ensuremath{{\scriptscriptstyle -}}}

\newcommand{\pspecind}{\ensuremath{\alpha}}
\newcommand{\turbind}{\ensuremath{\vka}}

\newcommand{\enhgas}{\eta_{ji}}
\newcommand{\enhdust}{\eta_{\rm\scriptscriptstyle dust}}
\newcommand{\enhmax}{\maxval{\eta}_{\rm\scriptscriptstyle gas}}

\newcommand{\diff}{\ensuremath{\mathrm{d}}}

\newcommand{\MeV}{\ensuremath{\mathrm{MeV}}}

\newcommand{\TeV}{\ensuremath{\mathrm{TeV}}}

\title{Thermal-to-nonthermal element abundances in different Galactic environments}
 \ShortTitle{Thermal-to-nonthermal element abundances}

\author*[a]{Bj\"{o}rn Eichmann}
\author[b]{J\"{o}rg P. Rachen}

\affiliation[a]{RAPP Center, Ruhr-Universit\"at Bochum, Institut f\"ur Theoretische Physik IV \\ 44780 Bochum, Germany}

\affiliation[b]{Vrije Universiteit Brussel (VUB), Astrophysical Institute \\ Pleinlaan 2, 1050 Elsene, Belgium}

\emailAdd{eiche@tp4.rub.de}
\emailAdd{jorg.paul.rachen@vub.be}

\abstract{The non-thermal source abundances of elements play a crucial role in the understanding of cosmic ray phenomena from a few GeV up to several tens of EeV. In this work a first systematic approach is presented that describes the change of the abundances from the thermal to the non-thermal state via non-linear diffusive shock acceleration by a temporally evolving shock. Hereby, not only time-dependent ionization states of elements contained in the ambient gas are considered, but also elements condensed on solid, charged dust grains, which not only can be injected into the acceleration process as well, but are from our findings even the dominant injection channel for most heavy elements. This generic parametrized model is then applied to the case of particle acceleration by supernova remnants in various ISM phases as well as Wolf-Rayet (WR) wind environments. We show that the overall low to medium energy cosmic ray distribution by WR explosions yield a significantly harder, which makes this contribution quite promising in order to explain the spectral hardening of the flux of certain elements, such as helium, observed by AMS-02 and other experiments at rigidities of about 1\,TV, which would also be an important test for the potential role of WR-progenitor supernovae as the sources of Galactic cosmic rays around the second knee. 
}

\FullConference{37$^{\rm{th}}$ International Cosmic Ray Conference (ICRC 2021)\\
		July 12th -- 23rd, 2021\\
		Online -- Berlin, Germany}

\begin{document}
\maketitle

\section{Introduction}


Diffusive shock acceleration (DSA) is still the prime candidate for the explanation of non-thermal particles, i.e. cosmic rays, at all energies. A key question is whether this process can deliver an understanding for the relation between the thermal element abundances in the ISM and the generally enhanced cosmic ray composition measured at ${\lesssim} 1\,$GeV (see, e.g., \cite{Caprioli+2014, Caprioli+2017, Hanusch+2019}). In a recent paper \cite{Eichmann+2021} (hereafter ER21), we presented a detailed discussion of the injection and acceleration of various ion species into the DSA process in the evolving shocks of supernova remnants, and found for the canonical warm neutral and warm ionised phases of the ISM excellent agreement with existing measurements low energy cosmic ray abundances. A key assumption to achieve this agreement for all measured elements up to zinc was the assumption that not only atomic ions, but also ionized dust grains are injected into the acceleration process \cite{EllisonDruryMeyer1997, MeyerDruryEllison1997} -- a mechanism that has been proposed more than 40 years ago as a solution to the so-called DSA injection problem \cite{Epstein1980}.  
Encouraged by this success, we now want to extend this model to higher energies, first to the energy range covered by the AMS-02 experiment ($100\,\MeV{-}1\,\TeV$) that recently delivered precise data on the spectra of various nuclear species from hydrogen up to iron \cite{Budrikis2021_AMS-02}. 
Here we encounter another problem to solve, which is
the cumulative non-thermal DSA spectra produced by supernova shocks during their evolution through various phases. The goal is to understand how the spectra of supernova remnants can match the injection index for cosmic rays in the Galaxy, $\alpha_{\rm inj}\approx 2.3$, required by cosmic ray transport calculations (see, e.g., \cite{Thoudam:2016syr,Thoudam+2021_icrc}), and to which degree the resulting cumulative injection spectra can be considered power laws at all. In this paper, we want to summarize some of the central aspects of ER21 and present first results on the correlated spectral behaviour in the model's parameter space.    

\section{Particle evolution from thermal-to-non-thermal energies}
Starting from a thermal pool of particles, whose macroscopic quantities, such as the temperature $T$ or density $n$ of its electrons and ions, respectively are determined from the given ISM phase of their habitation. 
We are interested in the abundance of various species of ions after they encountered a shock wave, which is known to redistribute them in power-law like non-thermal spectra by the mechanism of diffusive shock acceleration (DSA, see, e.g., \cite{Drury:1983zz}). 
The critical momentum $\pinj$ where supra-thermal particles can be injected into DSA has been a matter of debate for some time (e.g.\ \cite{Shapiro1997ICRC}, \cite{Caprioli+2014}). Recent results from 2D particle-in-cell (PIC) simulations \cite{Caprioli+2017, Hanusch+2019} indicate that $\pinj\simprop A$, so that we suppose that
\begin{equation}
 \pinj = \xinj A_j m_p \bsh c
\end{equation}
with the shock velocity $\bsh\,c$ and the parameter $\xinj\sim 3$ \cite{Caprioli2012}. Further, these simulations have shown that also the injection rate $\enhgas$ of gas phase elements (relative to protons) depends on the mass-to-charge ratio $A_j/Q_{ji}$ of the given gas phase species $j$ with an ionization state $i$. 
In addition, it needs to be taken into account that the particle is not immediately injected after passing the shock but resides a timescale of $\tscale[inj]{,ji}{} \propto \cdiff{\down}{}\!(\pinj)\propto (A_j/Q_{ji})^{2-\turbind}$ downstream of the shock before entering the acceleration process. Here \cdiff{\down}{} denotes the downstream diffusion coefficient, which is likely dominated by resonant scattering off magnetic turbulence generated by the cosmic-rays themselves (the so-called Bell-instability \cite{Bell2004}). Thus the spectral index of the turbulence spectrum is rather flat, i.e. $\turbind=1$, and $\tscale[inj]{,ji}{}\sim (0.1\dots10)\,\text{hr}$. Within this comparable short timescale the ionization fraction $\ionfrac{ji}$ of element species $j$ hardly changes from its initial values far upstream of the shock \cite{Eichmann+2021}.
Finally, the supra-thermal particles with $p\geq\pinj$ enter DSA and become non-thermal --- so-called cosmic-rays (CRs).
The differential number density of cosmic rays of species $j$ that arises at a time $t$ can be written as 
\begin{equation}
n_j(p,t)\equiv \frac{\diff n_j}{\diff p} = \frac{(\pspecind-1)\,\tilde{n}_j^{\langle \rm env \rangle}}{\pinj}\, \left(\frac{p}{\pinj}\right)^{-\pspecind}\;,
\label{eq:dndp}
\end{equation}
with the normalization constant $\tilde{n}_j=\Enhance{j} \,\density{}{}\, \abundance{j}$ that depends on the fraction $\Enhance{j}$ of the element species $j$ that crosses the shock and gets injected into DSA, the number density $\density{}{}$ of (charged \emph{and} neutral) atoms in the ambient shock environment, and the normalized thermal abundance $\abundance{j}$ of element species $j$ 
(we given an expression for $\Enhance{j}$ below). The power law index $\pspecind$ depends in a well-known way on the adiabatic index of the plasma and the Mach number of the shock. 
Here, we focus on the low-energy cosmic ray (LECR) part of the spectrum, i.e.\ at energies $\lesssim 1\,\text{TeV}$, so that we do not need to account for cut-off effects, which are expected to become relevant above rigidities of multiple TV. We suppose that 
$\pspecind$ does not depend on the element species
as expected in linear DSA theory,
though there are hints that for small Mach numbers, which are typically reached at the late Sedov-Taylor phase of the shock evolution (see below) deviations from this assumption might be possible \cite{Hanusch+2018}.
The spectra (\ref{eq:dndp}) are provided quasi-instantaneously at the shock and subsequently advect downstream or escape from the upstream --- whereof the latter only becomes significant at the high-energy end of the spectrum (e.g.~\cite{Caprioli2012}). Those CR spectra that reside within the downstream region until the SNR merges with its ambient environment suffer from adiabatic losses due to the expansion of the shock front with radius $r_{\rm sh}$. Hence, the overall differential number of cosmic rays of species $j$ that is ejected by an SNR at the end $t_{\rm f}$ of its lifetime is given by
\begin{equation}
N_j(p,\,t_{\rm f})\equiv \frac{\diff N_j}{\diff p} = \int_0^{t_{\rm f}} \diff t\,\, n_j(p,\,t)\,\area[sh]{}{}(t)\, \bsh(t)\,c\,\Lambda_{\rm ad}(t,\,t_{\rm f})\;.
\label{eq:dNdp}
\end{equation}
Here, \area[sh]{}{}(t) denotes the surface area of the shock at a given time and the impact of adiabatic losses on the particle distribution is given by (see e.g.~\cite{Dermer+2001})
\begin{equation}
    \Lambda_{\rm ad}(t,\,t_{\rm f})=\frac{r_{\rm sh}(t_{\rm f})}{r_{\rm sh}(t)}\,\left[\frac{\Gamma_{\rm sh}(t)}{\Gamma_{\rm sh}(t_{\rm f})}\right]^{1/3}
\label{eq:adiabLoss}
\end{equation}
with the Lorentz factor $\Gamma_{\rm sh}=[1-\bsh]^{-1/2}$. So, 
before we discuss the acceleration probability $\Prob{}{}$ in more detail we want to look at the impact of shock evolution 
on the differential overall particle number (\ref{eq:dNdp}).

\subsection{Shock evolution}
The acceleration of Galactic CRs is typically associated with DSA by a shock front that result from the interaction of stellar material ejected by a supernova with the ambient gas --- a so-called supernova remnant (SNR). First, the ejecta from the stellar progenitor are highly supersonic preceding a so-called blast-wave shock which compresses and heats the ISM environment. This stage is usually referred to the so-called \emph{free-expansion} (FE) (or \emph{ejecta-dominated}) phase, as the stellar material expands quasi-unperturbed. 
After a sufficient amount of material has been swept up to balance the mass of the ejected material,
the counteraction of the shocked ISM gas leads to the formation of a reverse shock, that initiates the so-called \emph{Sedov-Taylor} (ST) phase,
a pressure-driven adiabatic expansion phase 
in which
the shock speed decreases (typically according to $\bsh(t)\propto t^{-3/5}$) and the expansions decelerates.
This transition happens at a time $\tau_{\rm ST}$, which is at the order of several years up to a few hundreds of years dependent on the ambient medium. 
Efficient particle acceleration typically stops when the ejecta reaches a critical temperature, where free electrons get captured, so that radiative energy losses set in and the adiabatic expansion breaks down. This phase is usually referred to the so-called \emph{(pressure driven) snowplow} (PDS) phase.\footnote{We adopt the SNR evolution code \cite{Leahy+2017} to obtain a detailed description of the temporal development of an SNR in different media.} At these late phases of the SNR evolution, the SNR has typically reached quite low Mach numbers, so that the total compression ratio $R_{\rm tot}$ starts decreasing yielding steeper spectra than in its early phases.

The situation is somewhat different if we consider 
the case of a non-uniform ambient environment that is governed by 
an extended
stellar wind, such as for Wolf-Rayet (WR) stars
which are thought to be the progenitors of type 1b/c supernovae (\cite{Crowther2007}, henceforth referred to as WR-SN).
Here the shock can already reach the wind termination shock at about the time of transition into the ST phase. 
When passing this discontinuity, the rapid change in ambient pressure ca be assumed to rip apart the shock, 
or at least prohibit an efficient particle acceleration afterwards, so that low Mach number phases are never encountered. Additionally, 
the radiation-to-gas pressure ratio $P_{\rm rad}/P_{\rm gas}\gtrsim 1$
leads to an adiabatic index that 
is given by (e.g.~\cite{Turolla+1988})
\begin{equation}
    \adindex[WR]{}= 1 + \frac{2\,(1+4P_{\rm rad}/P_{\rm gas})}{3\,(1+8P_{\rm rad}/P_{\rm gas})}<5/3
\end{equation}
yielding a higher total compression ratio $R_{\rm tot}$ and consequently, a harder spectrum than in the standard case of non-relativistic gas, where $\adindex[ISM]{}=5/3$. 

\begin{figure}[tbp]
    \centering
    \includegraphics[width=0.56\linewidth]{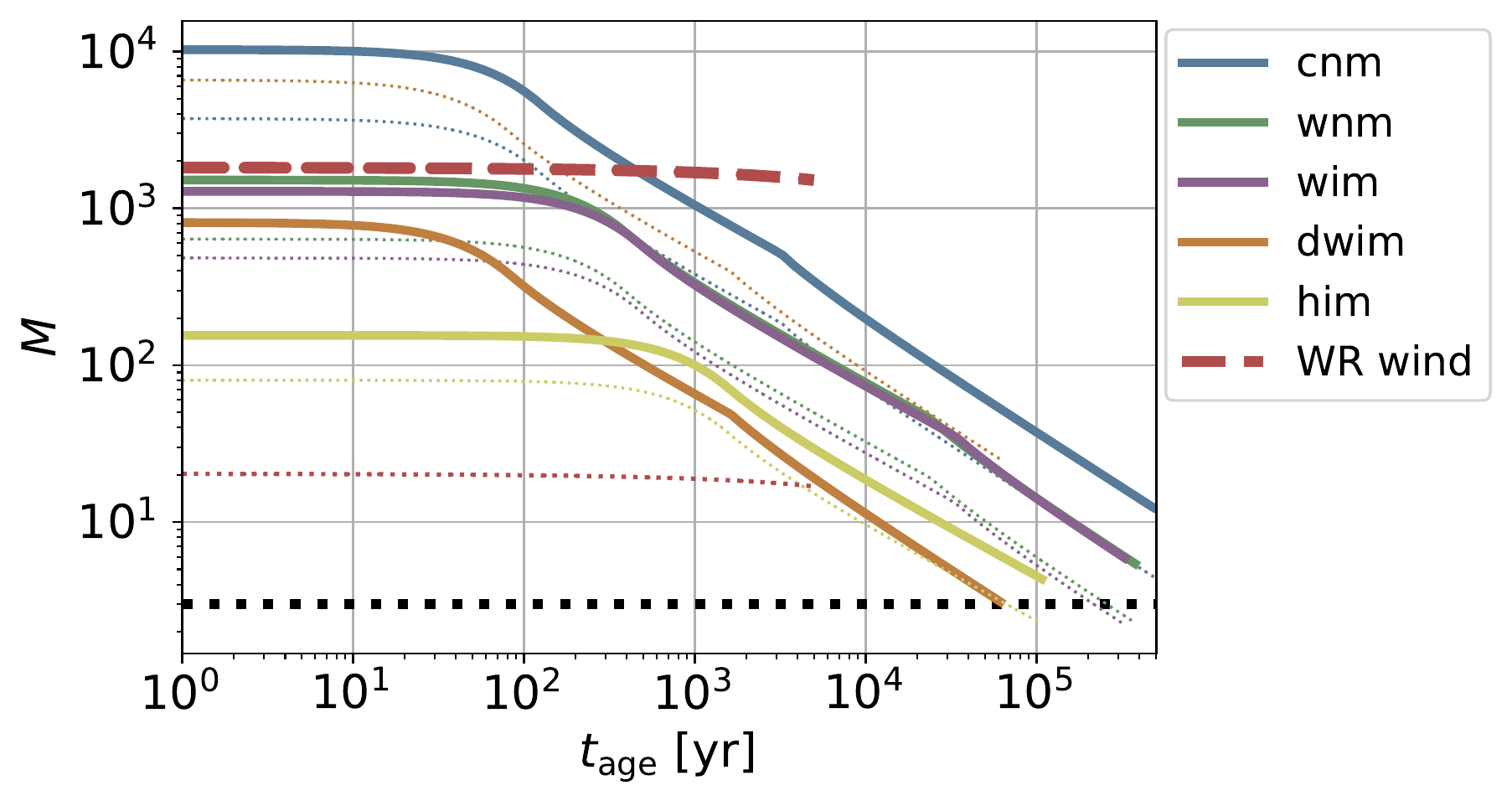}
    \includegraphics[width=0.43\linewidth]{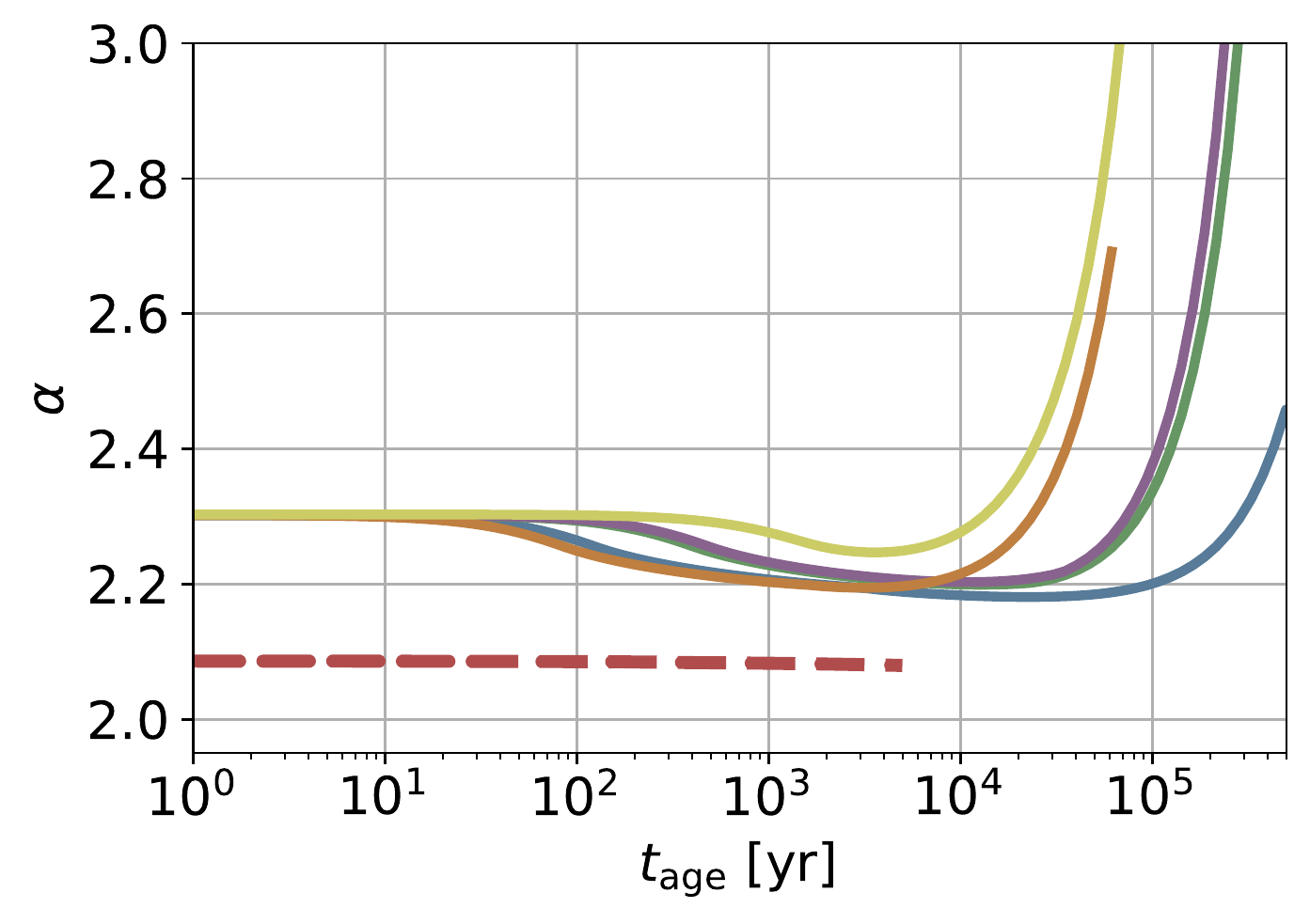}
    \caption{The temporal evolution of the Mach numbers (\emph{left}) --- thin lines indicate the Alvf\'{e}n Mach number and thick lines refer to the sonic Mach number --- and the spectral index (\emph{right}) dependent on the ambient shock environment.}
    \label{fig:shockEvol} 
\end{figure}

Including recent results on non-linear DSA \cite{Haggerty+2020, Caprioli+2020} we obtain a spectral behaviour that can become significantly steeper than the popular value of $\alpha=2$ from linear DSA at high Mach numbers
for the case of a constant ambient ISM environment. 
Two of the main parameters that fix the spectral index are the adiabatic index \adindex{} of the upstream gas as well as the normalized downstream magnetic and CR pressure, $\xi_{\rm B}$ and $\xi_{\rm CR}$, respectively. According to these non-linear effects the Fig.~\ref{fig:shockEvol} shows that the spectrum slightly flattens when the shock evolves into the SD phase (first "kink") before steepening in the PDS phase (second "kink") --- except for the cold neutral medium (CNM) where significant shock acceleration might be generally problematic (for more details see \cite{Eichmann+2021} and references therein), and the hot ionized medium (HIM). For the latter the SNR is still in the ST phase when the pressure of the shocked gas reaches the critical pressure of the ambient HIM gas, so that the shock breaks down. Supposing that $\xi_{\rm B}$ and $\xi_{\rm CR}$ do not depend on the ambient SNR environment, the uniform ISM environments generally show quite similar values of $\alpha$, which slightly depends on the charged particle density $\density{ion}{}$ of the ambient medium in the ST phase. But as the transition between the different evolution phases depends also on $\density{ion}{}$, the spectral index at a given SNR age differs significantly between the ISM phases. Due to the adiabatic losses (\ref{eq:adiabLoss}) as well as the increasing shock surface area \area[sh]{}{}(t), the late ST phases typically provide the dominant contribution to the overall differential number $N_j(p,\,t_{\rm f})$, hence, we expect that the minimal value of $\alpha$ yields the spectral behavior of $N_j(p\gg\pinj,\,t_{\rm f})$. 
Significantly flatter spectra without strong evolutions effects are found for the case supernovae in of WR winds.

\subsection{Total fraction of injected elements in gas and dust}
In order to determine the acceleration probability of heavy elements, we need to introduce another ISM phase not discussed so far: dust. Certain elements in the ISM (in particular its cold phases), such as Mg, Al and Ca, are predominantly bound in dust grains. During the acceleration process we suppose that dust grains behave like ions of a high mass-to-charge ratio, and that the sputtered grain material receives the same velocity as its parental grain \cite{Epstein1980, EllisonDruryMeyer1997}. In terms of the injection probability of dust grains there is even less knowledge, so that we adopt that $\Prob{\rm dust}{}=\Prob{0}{} \enhdust$, where 
\Prob{0}{} is the injection probability of a proton and
we condense all other dependencies on mass, charge or structure of the dust grain into a single parameter that we call the dust enhancement factor $\enhdust$. 

With respect to the gas phase elements, several simulations suggest a linear dependence of $\log\enhgas$ on $\log A_j/Q_{ij}$, until some maximum enhancement $\enhmax\gtrsim 10$ is reached \cite{Caprioli+2017,Hanusch+2018}. Different simulations disagree somewhat on the constant of proportionality, the details are discussed in ER21. It is there shown, however, that this detail, as well as other effects like ionization of gas elements by the shock, has little impact on the ability of fitting LECR data with our model, so that we eventually end up with two free parameters, $\enhmax$ and $\enhdust$ that can be adjusted to obtain correct enhancements for all elements, but which we believe are fundamental to DSA and thus should not depend on the external environment of a shock. 
Noting that for pre-shock ionization fractions $\ionfracUp{ji}$ the acceleration probability of a gas element $j$ can be approximated by
\begin{equation}
    \Prob{\rm gas,j}{\langle \rm env \rangle} \simeq \Prob{0}{}\,\sum_{i=1}^{Z_j} \enhgas\,\ionfracUp{ji}
\label{eq:neutrGasPro}
\end{equation}
we can determine the total fraction of injected elements from the environment-dependent normalized fraction $\fraction{j}$ of the elemental species $j$ that resides in the gas. Moreover, we suppose that everything that is not in the gas phase is in the dust phase, so that $\left(1-\fraction{j}\right)$ denotes the fraction elements that resides in dust. So, the only role of the dust grain is to shift the atoms bound in it across the threshold for supra-thermal injection before they start to get accelerated individually, which is why we expect no $\xinj$ dependence in Eq.~\ref{eq:dndp} for dust injection.
Eventually, the total fraction of injected elements is therefore given by the sum of gas and dust phase elements yielding
\begin{equation}
    \Enhance{j} = \fraction{j}\,\Prob{\rm gas,j}{\langle \rm env \rangle} + \left(1-\fraction{j}\right)\,\xinj^{1-\pspecind} \, \Prob{\rm dust}{}\,.
\end{equation}

\section{Predictions on the LECR spectra and extension to higher energies}
ER21 have already shown that an appropriate explanation of the (expected) LECR source abundances \cite{Voyager2016, ACE-CRIS2018} requires that a significant fraction of SNRs evolve into an ISM environment, that features ionization fractions $\ionfracUp{ji}$ and gas fractions $\fraction{j}$ as expected from the (dense) warm ionized medium (D)WIM. However, also SNRs in the warm neutral medium (WNM) can contribute significantly, if we account for the UV photons by the reverse fluorescence process that illuminate the shock site so that $\ionfracUp[WNM]{ji}\simeq\ionfracUp[WIM]{ji}$ (see \cite{Eichmann+2021} for a more detailed discussion). In contrast, the CNM as well as the HIM yield rather poor fits to the (indirectly) observed LECR source data, but still a subdominant contribution of the latter one cannot and should not be excluded. Further, these results show that a very efficient acceleration of dust grains, with $\enhdust\gg\enhgas$ is needed.

\begin{figure}[bp]
    \centering
    \includegraphics[width=0.75\linewidth]{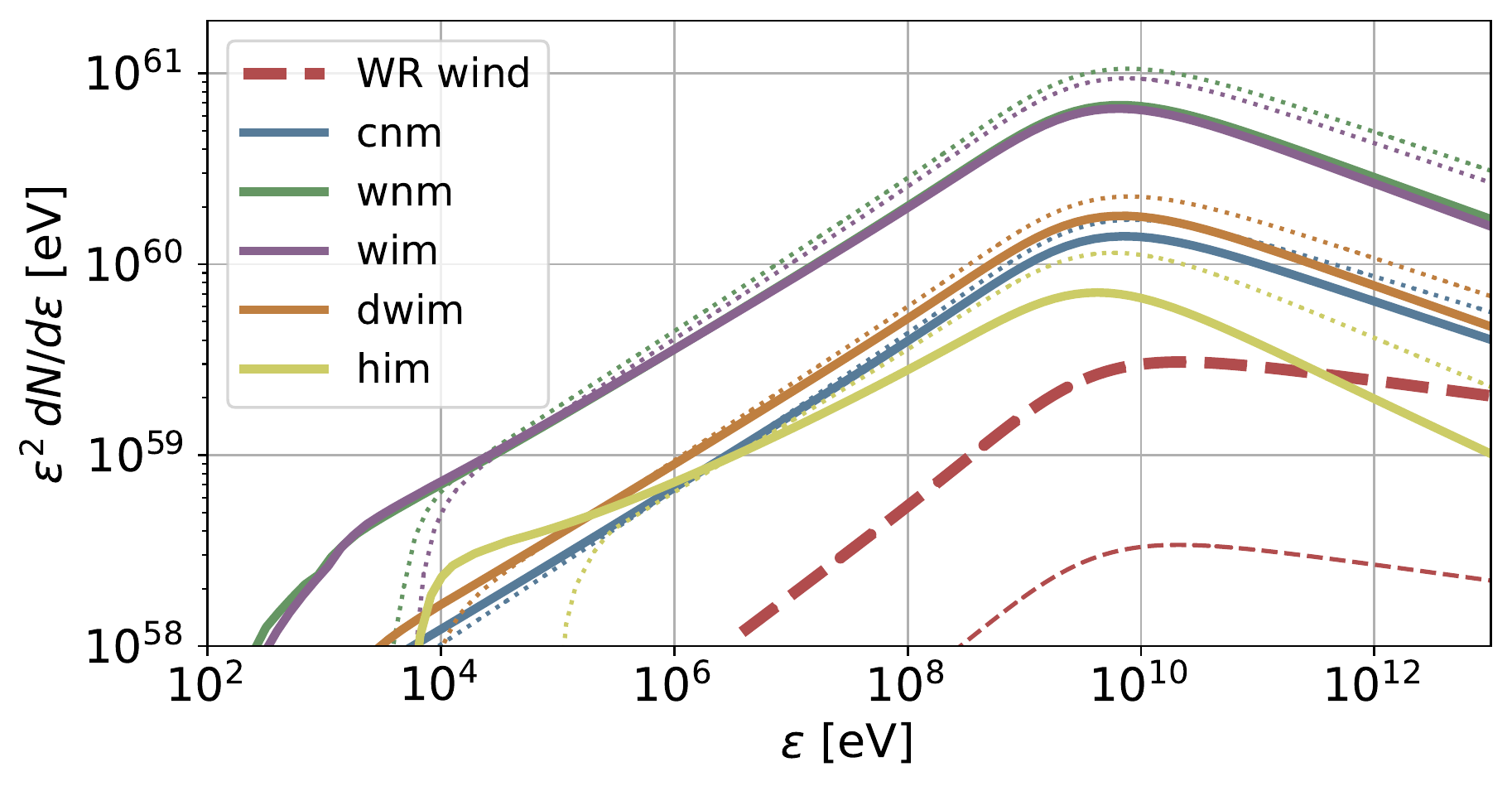}
    \caption{The overall differential number of helium nuclei at $t=t_{\rm f}$ (thick lines), as well as at $t=t_{\rm f}/10$ (thin lines). 
    \label{fig:spectrum} 
    }
\end{figure}

\begin{figure}[tbp]
    \centering
    \includegraphics[width=0.49\linewidth]{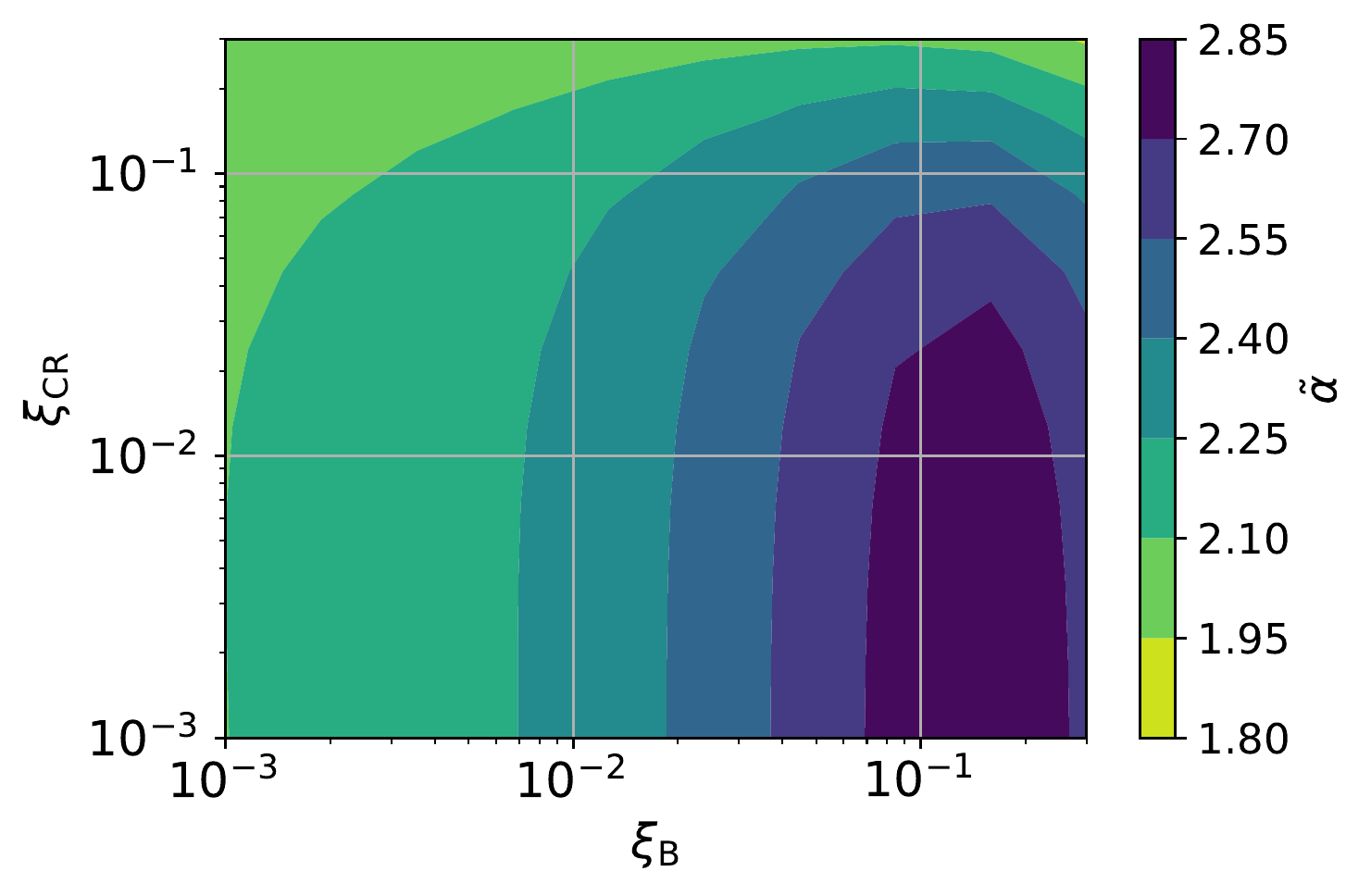}
    \includegraphics[width=0.49\linewidth]{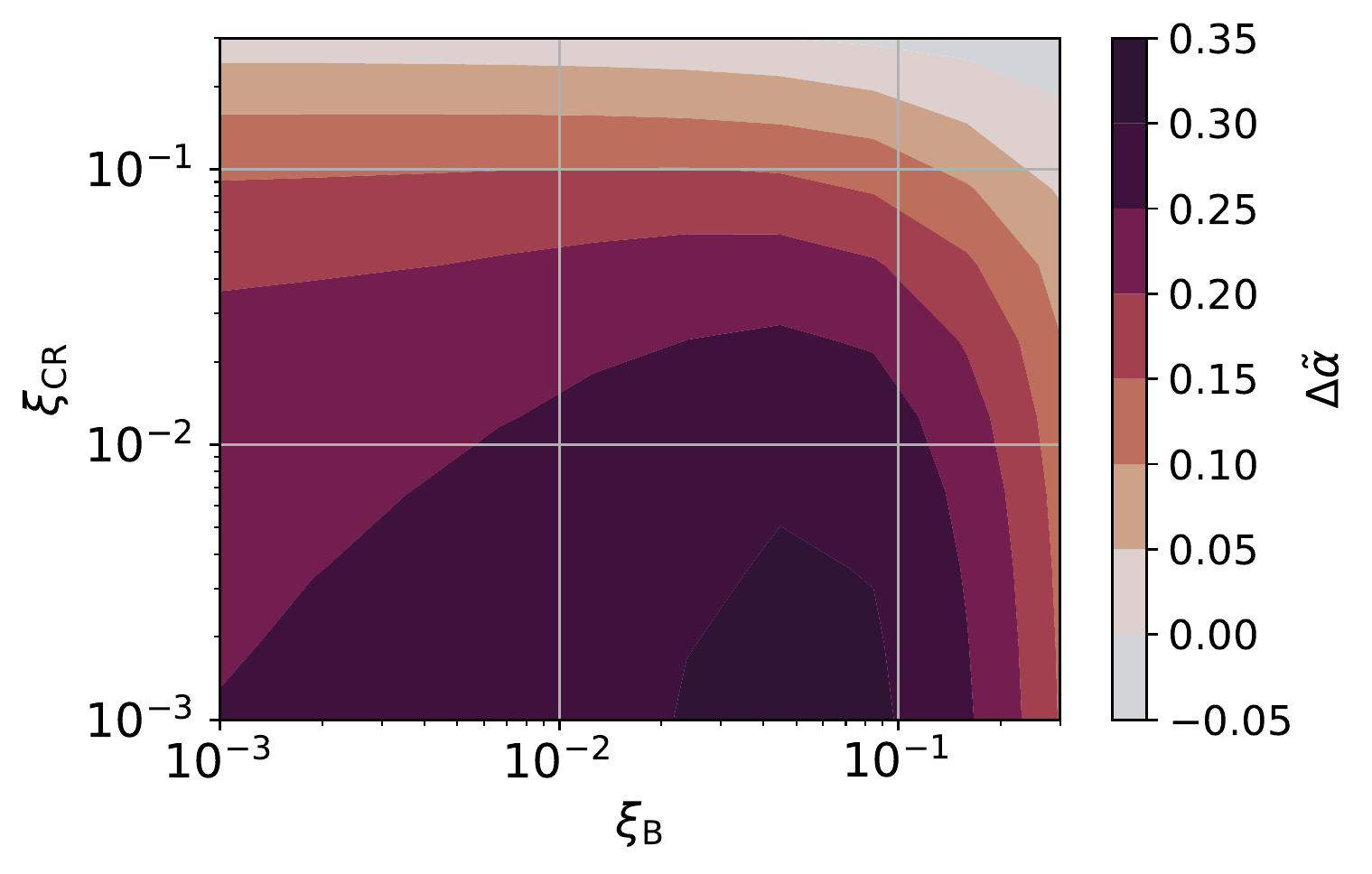}
    \caption{The overall spectral behavior between 2\,GeV and 1\,TeV dependent on $\xi_{\rm B}$ and $\xi_{\rm CR}$. \emph{Left:} The overall spectral index $\tilde{\alpha}^{\langle \rm WIM\rangle}$ that results from DSA in the WIM. \emph{Right:} The difference $\Delta \tilde{\alpha} = \tilde{\alpha}^{\langle \rm WIM\rangle}-\tilde{\alpha}^{\langle \rm WR\rangle}$ of the overall spectral index between the WIM and WR wind environment.
    \label{fig:specIdx} 
    }
\end{figure}

Beyond our investigation in ER21, we apply here our method also to WR-SN. These are considered a good candidate for the explanation of Galactic cosmic rays beyond the knee (see \cite{Thoudam:2016syr} and references therein), but they have never been considered a plausible source for LECR. We could show that this neglect is justified, as we obtain a very poor fit with a minimal reduced chi-squared value $\chi_\nu^2\gg 100$ for en exclusive WR-SN origin of LECR. To check how much WR-SN can contribute at higher ($\sim$\,TeV) energies, we need to discuss their overall energy distribution (\ref{eq:dNdp}) as a function of kinetic energy per nucleon  $\epsilon=\sqrt{(pc/A)^2+(m_pc^2)^2}-m_pc^2$, to that of regular SNR in different environments, as shown in Fig.~\ref{fig:spectrum}.
The normalized downstream magnetic and CR pressure, $\xi_{\rm B}$ and $\xi_{\rm CR}$, respectively, have no effect on the relative\footnote{The abundances are commonly normalized by the silicon abundance at a given kinetic energy per nucleon.} non-thermal element abundances, but a significant impact on the overall spectral index $\tilde{\alpha}$ as shown in the left Fig.~\ref{fig:specIdx}. In contrast, the enhancement parameters $\enhdust$ and $\enhgas$ steer the normalization of Eq.~\ref{eq:dNdp} of individual elements, but they have no impact on their spectral behaviour. Comparing the LECR spectra of helium, dependent on the given SNR environment, Fig.~\ref{fig:spectrum} exposes that at $\epsilon>m_{\rm p}c^2$ the uniform ISM environments, in particular the HIM, yield a significantly steeper spectral behaviour of $N_j(p,\,t_{\rm f})$ as expected. In addition, the long-lasting FE phase of the WR-SN leads to comparable large value of $\area[sh]{}{}(t_{\rm f})$, thus although the SNR age in these environments can hardly exceed a few kyr, WR wind environments are in principle able compete in terms of LECR contribution with regular SNRs in any other ISM environment. At about 1\,TeV this contribution might even exceed the one from SNRs in the HIM -- which captures up to 70\% of the Galactic volume -- for the unlikely case that SNRs are equally distributed in the different environments. According to the right panel in  Fig.~\ref{fig:specIdx}, non-linear DSA in WR wind environments yields a by $\Delta \tilde{\alpha}\sim 0.3$ flatter overall spectral index of $N_j(p,\,t_{\rm f})$, at energies of a few GeV for typical downstream magnetic and CR pressures of a few percent.

\section{Conclusions}
Recent progress in the decoding of the observed composition of CRs increases the need for a reliable physical model, that explains the change of abundances from the thermal to the non-thermal energy regime. Detailed knowledge on the observed energy distribution of individual element up to several hundreds of TeV allow conclusive predictions on the necessary LECR source abundances. We presented a model that (a) considers the detailed ionization fractions of gas-elements in the shock environment, (b) allows for the injection of refractory elements in the acceleration process via charged dust grains, (c) accounts for the temporal evolution of the shock, and (d) includes recent findings on the non-linear effects of DSA as well as the mass-to-charge dependent enhancement of gas-elements at the injection process. 
Based on the fitting procedure that has been introduced in a recent work \cite{Eichmann+2021}, we could verify that supernova explosions in the (D)WIM still provide the best agreement to the (indirectly) observed LECR source abundances. Including a reasonable mixture of SNR environments and optimizing the few model parameters we have, it seems feasible to explain the observed features of LECR spectra using our model predictions. Supernovae from Wolf-Rayet progenitors play no significant role at $\sim$\,GeV LECR, but due their harder energy distribution of cosmic rays owing to fast shocks in an environment of high radiative pressure, they can have a measurable impact already at $\sim$\,TeV energies. Because of the quite different thermal abundances in WR winds \cite[see references in][]{Thoudam:2016syr} they may leave there a distinctive signature for certain elements, which in comparison to high-precision data as provided by the AMS-02 experiment could be used to test their potential role in the explanation of cosmic rays around the second knee \cite{Thoudam+2021_icrc}.


\bibliographystyle{JHEP}
\addcontentsline{toc}{section}{Bibliography}
\bibliography{references}


%

%
%
%

\end{document}